\documentstyle[amssymb,aps,prb,floats,twocolumn,epsfig]{revtex}

\begin{document}

\wideabs{
\title{Free carrier effects in gallium nitride
epilayers: \\ the valence band dispersion.}

\author{P. A. Shields, R. J. Nicholas\cite{robin}} \address{Department of Physics,
Oxford University, Clarendon Laboratory, Parks Rd., Oxford, OX1 3PU, United
Kingdom}
\author{F. M. Peeters\cite{francois}}
\address{Department of Physics,
University of Antwerp, Universiteitsplein 1, B-2610 Antwerpen, Belgium.}
\author{B. Beaumont, P. Gibart} \address{CNRS, Centre
de Recherche sur l'H$\acute e$t$\acute e$ro-Epitaxie et ses Applications,
Valbonne, F-06560, France.}
\date{\today }
\maketitle

\begin{abstract}

The dispersion of the A-valence-band in GaN has been deduced from the
observation of high-index magneto-excitonic states in polarised interband
magneto-reflectivity and is found to be strongly non-parabolic with a mass in
the range 1.2--1.8$\,m_{e}$. It matches the theory of Kim et al.
\symbol{91}Phys. Rev. B 56, 7363 (1997)\symbol{93} extremely well, which also
gives a strong $k$-dependent A-valence-band mass. A strong phonon coupling
leads to quenching of the observed transitions at an LO-phonon energy above the
band gap and a strong non-parabolicity. The valence band was deduced from
subtracting from the reduced dispersion the electron contribution with a model
that includes a full treatment of the electron-phonon interaction.
\end{abstract}
\pacs{78.20.Ls
78.40.Fy
71.35.Ji
71.38.Fp
 } }
\narrowtext

Cyclotron resonance experiments have yet to shed significant light on the
valence band structure of gallium nitride,\cite{Puhlmann} despite using
magnetic fields up to 700 T. Therefore less direct methods are required to
experimentally determine the effective mass parameters. Several techniques have
been employed (see Ref. \onlinecite{Kasic} for review), but the significant
scatter of the resulting hole masses varying from 0.5-2.2 $m_{e}$ illustrates
their limitations. In contrast, the magneto-optics of GaN in the last few years
has been successful in providing an accurate mass for the conduction
band.\cite{Witowski,Drechsler}

Interband magneto-optics has made sig\-nificant pro\-gress recently as a result
of the improvements in the luminescence linewidths in the bandgap region.
Beyond successfully examining the impurity bound states, so far the analysis
has centred on the properties of the strongly bound excitonic states such as
the $1s$ and $2s$ free excitons. The properties of these are considerably
complicated by the proximity of the split valence bands where it has been
pointed out that a full description of the magnetic field dependence of these
states requires account to be taken of the interaction of all the excitonic
states belonging to the different valence bands. \cite{Rodina01}

In order to find out more about the valence band and to avoid this problem, we
present a study of magneto-reflectivity data from high order Landau level
transitions that have energies that are considerably greater than the excitonic
binding energy. By looking at the high index levels the magneto-excitonic
corrections become less significant, so that the transitions are dominated by
the contributions from the free carriers. Then, by considering in detail the
properties of the conduction band, now accurately known to have a mass of
$0.2220(5)$ $m_e$,\cite{Witowski} the valence band properties can be deduced
with reasonable confidence.

\medskip

The polarised magneto-reflectivity experiments were performed in a continuous
magnetic field up to 20 T and at 4.2 K. The sample, G889, was grown by the ELOG
method by MOVPE on a sapphire substrate.\cite{Beaumont} The hexagonal c-axis
was parallel to the magnetic field and perpendicular to the electric field
vector of the light. Fibre optics were used to guide the 75W Xenon light source
to the sample, also taking the reflected light to a quarter metre spectrometer
fitted with a UV-enhanced CCD detector.

\medskip

\begin{figure}
\epsfxsize=85mm \epsffile{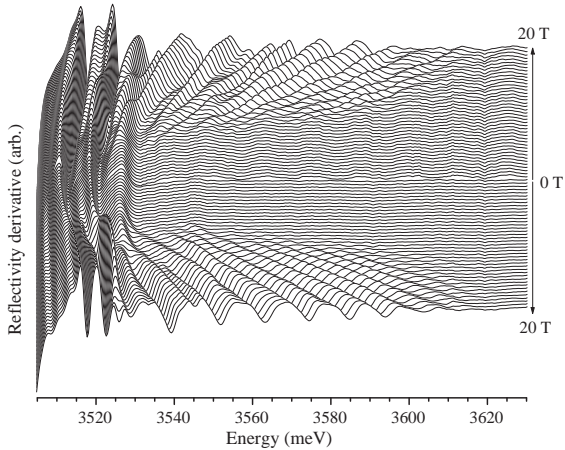} \caption{The derivative of the
reflectivity, $-dR/dE$, is shown for the two circular polarisations for
magnetic fields up to 20 T. The energy range includes the excited $n=2$ exciton
states as well as the `free carrier' Landau levels. The transitions are
quenched around 3620 meV, as a result of possible LO phonon emission above this
energy. $\sigma^+$: top, $\sigma^-$: bottom.} \label{fig1}
\end{figure}

The reflectivity spectra were analysed by identifying peaks in the first
derivative, $-dR/dE$\cite{Stepniewski19996,Shields99} to give the transition
energies, as shown in Figure \ref{fig1}. This approximates to a Kramers-Kronig
analysis for the case of weak, well separated transitions and has been already
successfully applied to describe the Zeeman shifts of the 1s excitonic
states.\cite{Stepniewski19996}

The data for the two different polarizations were analysed separately and it
was found that for all of the higher index transitions it was possible to take
account of the Zeeman splitting by using a single value of the g-factor. A term
$\pm\frac{1}{2}g_s\mu_{B}B$ was subtracted with $g_s=-1.8(2)$, which was then
found to give co-incident results for the two polarizations. For comparison
this is significantly different to that determined for the A $1s$ excitonic
states, where we find the weak field g-factor is $g_s=+0.254$. Using a value of
$g_e=1.95$ for the electron g-factor \cite{Carlos,electronGfactor} gives for
the 1s exciton $g_{A}^{\parallel}=2.2(2)$, in agreement with Ref.
\onlinecite{Rodina01}, but which is then quenched to the much lower value of
$g_{A}^{\parallel}=0.2(2)$ for the higher transitions. The large difference in
the g-factors for the different states confirms the conjecture by Rodina
\emph{et al.}\cite{Rodina01} that there are large differences between the
g-factors of the holes when they are involved in different states, as a result
of inter-valence band coupling.

In addition to the Zeeman energy there is also the small Coulomb correction due
to the finite binding of the higher excitonic states that causes the
transitions to lie just below the corresponding Landau level energies.
Unfortunately there are no numerical values available for the higher level
states in the field range used here so it is necessary to make some
approximations. The binding energies are usually expressed as a function of the
dimensionless parameter $\gamma=\hbar\omega/2R$. For InSb it was found that for
the high field regime, $\gamma>5$, the binding energy was described accurately
by, \cite{Tang,Weiler,Johnson}
\begin{equation}\label{coulombcorrection}
  E_{ex} \simeq \lambda R \left[ \frac{\gamma}{2n+1} \right]^{1/3}
\end{equation}
where the prefactor $\lambda=1.6$. For GaN $\gamma=0.11\rightarrow0.27$ for
$B=8\rightarrow20$ T. Despite these small values of $\gamma$ for the 1s state,
we still expect the excited states to be dominated by the magnetic energy in
the high field regime, as a result of the reduced effective binding energy for
higher Landau levels.  We therefore adopt the functional form of Eq.
\ref{coulombcorrection} but adjust the value of the prefactor $\lambda$.
Numerical calculations do exist for a Landau level index of
n=3,\cite{Makado19869} so by comparing Eq. \ref{coulombcorrection} with a
weighted average of the multiple bound states associated with any one Landau
level, we expect a prefactor of $\gamma\sim0.5$.

Once stripped of the Landau level quantisation and the Coulomb corrections that
depend on this, one expects the transition energies from all LL indices to
reduce to a single simple \emph{E--k} dispersion relation. This provides us
with an empirical method of fine-tuning the Coulomb correction by optimising
the co-linearity of the different Landau index contributions to the dispersion
relations. We find that the optimum results are obtained with a value of
$\gamma=0.4$ which gives a typical binding energy for the higher levels of
order 2-3 meV.

In order to analyse the transition energies in terms of a single free carrier $E(k)$
dispersion we use a semi-classical quantization of the Landau levels with,
\begin{equation}\label{plasmaapprox}
  \langle k^{2} \rangle = eB(2n+1)/\hbar.
\end{equation}
This requires the assumption that the Landau quantisation is the dominant
effect, which is easily satisfied for the magnetic fields used.

\begin{figure}
\epsfxsize=85mm \epsffile{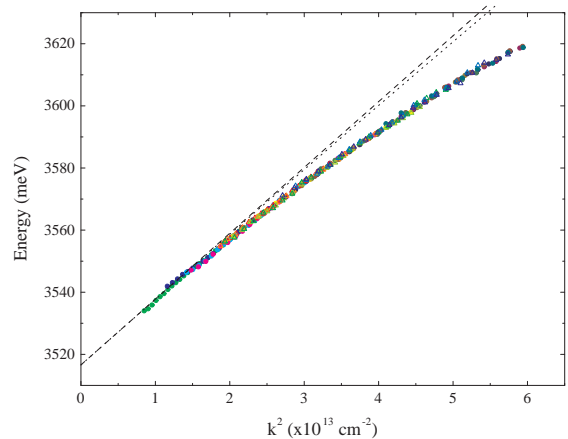} \caption{The reduced dispersion formed from
free carrier Landau levels. For polarisation $\sigma^-$ (closed circles) LL
indices n=3-13 were used, whereas for polarisation $\sigma^+$ (open circles)
n=5-10. Reduced dispersion without polaron corrections are shown (dashed line),
and including $k.p$ induced band non-parabolicity (dotted line). } \label{fig2}
\end{figure}

The reduced dispersion determined from the LL transitions is shown in Figure
\ref{fig2}. In order to reduce the significance of the excitonic Coulomb
interactions we have excluded the lowest transitions, leaving LL indices
\mbox{$n=$ 3--13} for $\sigma^-$, along with \mbox{$n=$ 5--10} for $\sigma^+$.

The extrapolation of the reduced dispersion to $k=0$ indicates that the Landau
levels are derived from the A valence band. The Rydberg energy was estimated
from the $1s$--$2s$ exciton separation in the same sample as used in this
work\cite{Neu199910} using a hydrogenic model.  Then from the calculated $2s$
binding energy, the band edge can be known to within $\pm1$meV, despite not
including the polaron corrections to the binding energy.\cite{Rodina01} No
Landau levels can be seen from the B band, but its mass is expected to be
considerably smaller, $m_B^{\perp}\sim0.35$.\cite{Kim19978} This will give both
a reduced oscillator strength and larger cyclotron splittings that will be
difficult to see underneath the strong A band transitions.

\begin{figure}
\epsfxsize=85mm \epsffile{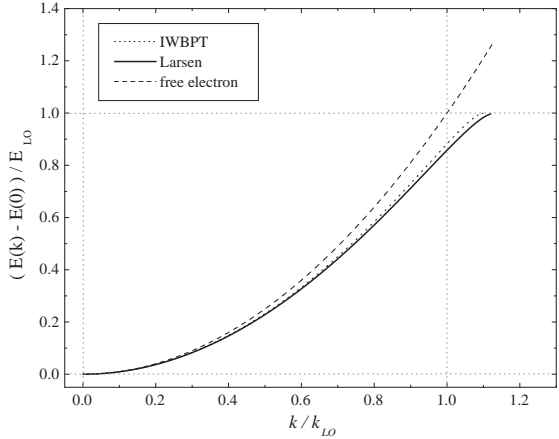} \caption{The polaron energy spectrum
calculated with perturbation theory (IWBPT) and variational techniques
(Larsen), for $\alpha=0.49$, compared with the free electron. } \label{fig3}
\end{figure}

The reduced dispersion is strongly non-parabolic, even after including the band
contribution using Kane's k.p framework in the two-band approximation,
\cite{Kane}
\begin{equation}\label{twoband}
  E-E_{g}=\frac{\hbar^2k^2}{2m^*}\left(1-\frac{1}{Eg}\frac{\hbar^2k^2}{2m^*}\left(\frac{m^*}{m_0}-1\right)^2\right).
\end{equation}
The additional non-parabolicity is due to the importance of resonant polaron
coupling in this material which is strongly polar.\cite{Wu} Further evidence
for this idea comes from the rapid disappearance of observable transitions at
an energy of $3620$ meV which is $\sim
\left(1+\frac{m_e}{m_h}\right)E_{LO}+E_g$ and is where resonant coupling will
occur for the electrons. This resonant polaron effect has a strong influence on
the electron energy-momentum relation, $E(k)$, particularly when the LO--phonon
energy is approached. A calculation of the electron--polaron dispersion
relation is shown in figure \ref{fig3}.

The electron-phonon coupling constant in GaN is sufficiently large,\cite{Wu}
i.e. $\alpha = 0.49$, that perturbation theory is no longer valid. In order to
calculate polaron effects, and in particular the energy-momentum
relation,\cite{Warmenbol} one has to go beyond the widely used `improved
Wigner-Brillouin perturbation theory' (IWBPT). A suitable theory, which is
valid for the intermediate electron-phonon coupling regime ($\alpha\leq1$), is
provided by the variational ansatz approach of Larsen\cite{Larsen} which is a
combination of the one-phonon Tamm-Dancoff approximation\cite{whitfield} and
the Lee-Low-Pines transformation.\cite{lee} The polaron energy-momentum
relation is obtained from a solution of the secular equation\cite{Larsen}

$$
\Delta E = k^2 + \left( 1 + \frac{\Delta E - k^2}{2k^2} \right)
A(\Delta E) \ ,
$$
with
$$
A = \frac{\alpha}{2\pi k} \left( \pi (1+c)k - \int^1_0 dx
\frac{1}{x} (D(c,x,p) + D(c,1/x,p)\right) \ ,
$$
where $D(c,x,p) = ((c+x^2)/(1+x^2))^2 \log((c+x^2+2xp)/(c+x^2-2xp))$ with $c=1
+ k^2 - \Delta E$. The above energy is in units of $\hbar \omega_{LO}$, the
LO-phonon energy, and the electron wavevector is in units of $k_{LO} =
\sqrt{2m\omega_{LO}/\hbar}$ where $m$ is the electron band mass. In the limit
of small polaron momentum one finds $\Delta E = E(k) - E(0) =
k^2((1-\alpha/12)/(1+\alpha/2))$ and consequently the small momentum polaron
mass becomes $m^*/m = (1+\alpha/12)/(1-\alpha/12)$ which equals $m^*/m=1.085$
for GaN ($\alpha =0.49$).

The polaron correction to the energy--momentum relation is two-fold. First
there is a renormalization of the electron mass into the polaron mass, $m^*$,
for small momentum/energy. Second, for larger energies close to the LO-phonon
energy, a resonant interaction occurs between the electron and the crystal,
resulting in a flattening of the $E(k)$ relation which leads ultimately to
$\partial E/\partial k =0$ at some critical $k^* > k_{LO}$. This resonant
polaron induced non-parabolicity correction to the $E(k)$ relation is
particularly important as the LO phonon energy is approached as is apparent in
both experiment (Figure \ref{fig2}) and theory (Figure \ref{fig3}). In
accounting for this effect, we use the value for the low-energy polaron mass of
\mbox{$0.2220(5)$$m_e$}, \cite{Witowski} from which we deduce a bare band edge
mass of \mbox{$m=0.204$ $m_e$}.

\begin{figure}
\epsfxsize=85mm \epsffile{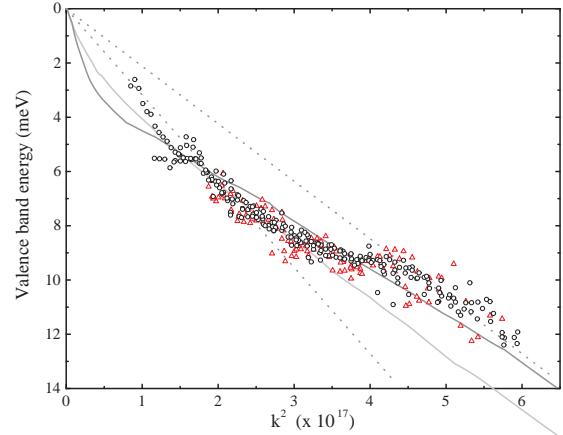} \caption{The valence band dispersion of the
A band deduced from the reduced dispersion with the electron contribution
determined by the Larsen approach. Solid lines show a theoretical calculation
for $k_{\pm}^{\perp}$ that includes the spin-orbit interaction \cite{Kim19978}.
Dashed lines show constant masses of $m_h=1.2$ and $1.8$. Black circles:
$\sigma^-$, Red triangles: $\sigma^+$} \label{fig4}
\end{figure}

Having taken proper account of the Coulomb and electron-polaron corrections to
the dispersion relation we can now deduce the valence band dispersion. This is
shown in Figure \ref{fig4} as a function of $k^{2}$. The data shows that the
valence band is considerably non-parabolic as can be seen by the straight lines
corresponding to parabolic masses of 1.2 and 1.8 in Figure \ref{fig4}. However
theoretically this is expected as a result of the spin-orbit
interaction,\cite{Kim19978} which splits the bands and leads to a much lighter
mass at $k=0$ and strongly energy-dependent hole masses. The results from this
theory,\cite{fig7a} for a slightly different value of the biaxial strain (and
hence A-B separation), are also shown in Figure \ref{fig4}. The agreement
between the experimental data points and the theory is quite good, although it
is not possible to access the very light mass region predicted by the theory
very close to the band edge.

\begin{figure}
\epsfxsize=85mm \epsffile{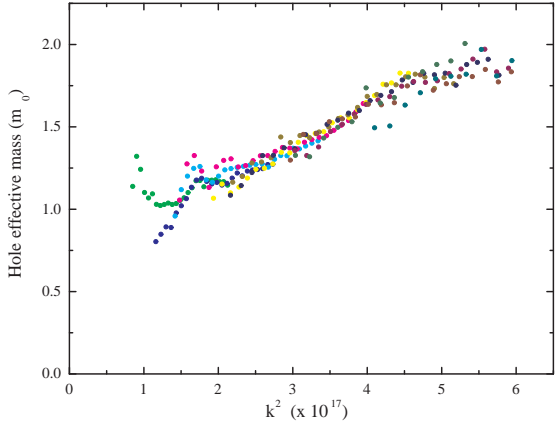} \caption{The k--dependent effective A-hole
mass for the $\sigma^-$ polarisation.} \label{fig5}
\end{figure}

In order to determine the k--dependent effective masses, we define the
quantity, $m(k)=k^2/E_k$, and these results are shown in Figure \ref{fig5}. The
effective masses can be seen to increase approximately linearly with $k^{2}$
from \mbox{$\sim$1.2--1.8 $m_{e}$}, suggesting a band edge mass value of
$m_h\simeq$ 0.8, although we cannot exclude a value lower than this in the
range $E\lesssim$ 5 meV. When comparing these values with theory, however, it
should be remembered that these values are measured at low frequency and will
therefore represent the `hole' values with additional polaron dressing from the
valence band and will therefore be somewhat higher than the bare band edge
values.

\medskip

To conclude, the results described show clearly the quite large mass values and
strong non-parabolicity of the A-valence-band in the direction perpendicular to
the c-axis. Consideration of the spin-orbit interaction is essential to
understand this effect. It has been predicted that the band is light at the
band edge\cite{Kim19978,Rodina01}, however our results show that this no longer
holds for energies $\gtrsim$ 5 meV into the band.

\medskip

This work is supported by the EPSRC (UK), Flemish Science Foundation (FWO-Vl),
IUAP-IV, the `Bijzonder onderzoeksfonds van de Universiteit Antwerpen', and the
British-Flemish Academic Research Collaboration Programme. P.A.S. acknowledges
financial support from Sharp Laboratories of Europe Ltd.


\begin{thebibliography}{000}
\bibitem{electronGfactor} The A valence band does not split
for  $B \perp c$, giving the electron g value in the free exciton. Our value of
$2.00\pm0.05$ agrees with the value 1.95 deduced from electron spin resonance.
\cite{Carlos}
\end{thebibliography}
\end{document}